\nonstopmode
\documentclass{aa}
\usepackage{xspace}
\usepackage{natbib}
\usepackage{graphicx}
\usepackage{txfonts}
\usepackage{aasjournals}

\bibpunct{(}{)}{;}{a}{}{,}

\newcommand{\lstar}{\ensuremath{\mathit{L}_{\star}}}                
\newcommand{\lsun}{\ensuremath{\mathit{L}_{\odot}}}                 
\newcommand{\mdust}{\ensuremath{\mathit{M}_{\rm dust}}}             
\newcommand{\micron}{\ensuremath{\mu {\rm m}}}                      
\newcommand{\um}{\ensuremath{\mu\mbox{m}}\xspace}                   

\newcommand{\mstar}{\ensuremath{\mathit{M}_{\star}}}                
\newcommand{\msun}{\ensuremath{\mathit{M}_{\odot}}}                 
\newcommand{\rdisk}{\ensuremath{\mathit{R}_{\rm disk}}}             
\newcommand{\rstar}{\ensuremath{\mathit{R}_{\star}}}                
\newcommand{\rsun}{\ensuremath{\mathit{R}_{\odot}}}                 
\newcommand{\teff}{\ensuremath{\mathit{T}_{\rm eff}}}               %
\newcommand{\tsub}{\ensuremath{\mathit{T}_{\rm sub}}}               
\newcommand{\degree}{\ensuremath{^\circ}}                           

\begin{document}

\title{A parameter study of self-consistent disk models around Herbig~AeBe~stars}

\subtitle{}
\author{J. Meijer\inst{1}
  \and  C. Dominik\inst{1,2}
  \and  A. de Koter\inst{1}
  \and  C.P. Dullemond\inst{3}
  \and  R. van Boekel\inst{3}
  \and  L.B.F.M. Waters\inst{1,4}}
\offprints{J. Meijer,\\
  \email{jmeijer@science.uva.nl}}

\institute{Sterrenkundig Instituut ``Anton Pannekoek'',
           Kruislaan 403,
           1098 SJ Amsterdam, 
           The Netherlands
\and       Department of Astrophysics,
           Radboud University Nijmegen,
           PO Box 9010,
           6500 GL Nijmegen,
           The Netherlands
\and       Leader of Junior Research Group "Formation of
           planetary building blocks" at the
           Max-Planck-Institut f\"ur Astronomie, K\"onigstuhl 17,
           69117 Heidelberg, Germany
\and       Instituut voor Sterrenkunde,
           Katholieke Universiteit Leuven,
           Celestijnenlaan 200D,
           3001 Heverlee, Belgium}

\date{DRAFT, \today}

\abstract{We present a parameter study of self-consistent models of
  protoplanetary disks around Herbig AeBe stars.  We use the code
  developed by Dullemond and Dominik, which solves the 2D radiative
  transfer problem including an iteration for the vertical hydrostatic
  structure of the disk.  This grid of models will be used for several
  studies on disk emission and mineralogy in followup papers.  In this
  paper we take a first look on the new models, compare them with
  previous modeling attempts and focus on the effects of various
  parameters on the overall structure of the SED that leads to the
  classification of Herbig AeBe stars into two groups , with a flaring
  (group I) or self-shadowed (group II) SED.  We find that the
  parameter of overriding importance to the SED is the total mass in
  grains smaller than 25\um, confirming the earlier results by
  Dullemond and Dominik.  All other parameters studied have only
  minor influences, and will alter the SED type only in borderline cases.
  We find that there is no natural dichotomy between group I and
    II.  From a modeling point of view, the transition is a continuous
    function of the small dust mass.  We also show that moderate
  grain growth produces spectra with weak or no 10\um feature, both
  for flaring (Group I) and non-flaring (Group II) sources.  The fact
  that sources with weak features have been found mostly in Group I
  sources is therefore surprising and must be due to observational
  biases or evolutionary effects.  }

\maketitle


\section{Introduction}

Herbig Ae/Be stars are young, intermediate-mass stars surrounded by
circumstellar disks (see e.g. \citet{ww98} or
\citet{2000prpl.conf..559N} for a review).  Solid-state material makes
up about a percent of the disk mass, and can be studied in the thermal
infrared.  These studies indicate that the grains must have
experienced both growth and thermal processing, since their spectra
differ substantially from that of interstellar dust
\citep{bmk01,bwd03,bmw05}. It is expected that this growth may lead
ultimately to the formation of planets. The infrared spectra of Herbig
stars can be divided into two groups \citep{mwb01}.  Though
approximately similar in the near-IR, the far-IR flux of some is
relatively strong (those of Group I) compared with that of others
(Group II). This spectral difference is believed to reflect the
overall shape of the disk.  Group I sources have a geometrically
flaring outer disk, which absorbs direct stellar light that is
thenre-radiated at far-IR wavelengths. Group II sources have a flat
geometry, which reprocesses far less stellar radiation in the outer
disk.  Many sources exhibit 10 micron silicate emission (these are
labeled $a$, so either Ia or IIa).  Some sources lack this emission
(these are denoted Ib or IIb), possibly reflecting that the bulk of
the observable warm silicate grains in these sources have sizes larger
than several microns.

The general shape of the SED of Herbig stars has been a subject of
study for a considerable time and is now believed to be quite well
understood.  The latest addition to this understanding was the
realization that the prominent emission close to 3\um, interpreted in
earlier models interpreted as originating in accretion luminosity
\citep[e.g.][]{1992ApJ...397..613H}, is due to a puffed-up inner rim
of the disk.  This rim is formed at a location given by the
dust-evaporation temperature. Inside the rim, the disk does not
contain any dust and is mostly transparent (optically thin) to
the incident stellar light.  The rim itself can absorb up to 25\% of
the stellar radiation \citep{2001A&A...371..186N}.  This idea was
incorporated into a more detailed physical model by
\citet{2001ApJ...560..957D}, who showed that this inner rim can indeed
resolve the discrepancy between observations and previous models in
the wavelength region around 3\um, with a self-consistent disk model.
An important question is if this conclusion still holds in more
complex models.

\citet{d02} and \citet[hereafter DD04]{dd04} developed far more
sophisticated models including full 2D radiative transfer, coupled
with self-consistent vertical structure calculations.  These models
confirmed that a self-consistent approach can explain many of the
observed features of Herbig Ae SEDs.  They demonstrated that the disk
structure is closely linked to the properties and spatial distribution
of the dust grains. Although the vertical extent of the disk is
supported by gas pressure, the energy that heats the gas is obtained
from the grains by dust-gas interactions (see \citet{kd04} for
details). The spatial distribution of {\em dust opacity} that
  absorbs stellar light must therefore control, to a large extent,
the geometrical shape of proto-planetary disks.  Since this opacity is
provided mostly by \textit{grains smaller than about 25\um} -- that
have large mass-extinction coefficients --  this grain
population is the mainly relevant one.  Using this principle, DD04 performed a
pilot study in which they showed that by increasing the total mass of
the small grain component, one can alter the overall appearance of the
spectral energy distribution from one qualitatively characteristic of
Group II into Group I, and that this corresponds to either a
non-flaring vs. a flaring geometry.  A change in the geometry could also
be achieved by a very steep power law in the surface density, which in
effect concentrates most of the disk mass in the inner rim,
  leaving the outer disk with very low mass and, consequently,
  opacity.

This paper is the first in a series where we explore these new models
in greater detail, building on the work of DD04. In this first paper,
we focus on the general appearance of the disk spectra, and how it
depends upon the various parameters entering the calculation.  We
present an extended parameter study of dust and disk properties of
HAeBe star with the aim of better quantifying and understanding the
cause of the Group I/II dichotomy more robustly as well as the
presence or absence of the 10\micron\ feature.  We do not consider the
detailed shape of the 10\um feature - mineralogical studies will be
presented in further papers in this series.

The model parameters that we address are -- in addition to total disk
mass and radial surface density gradient -- disk size and inclination,
some basic aspects of grain composition, and the grain-size
distribution.  To allow for a {\em quantitative} comparison between
observations and models, we present the results in terms of a
color-flux ratio comparison, first introduced by \citet[hereafter
BWD03]{bwd03} to distinguish between Group I and Group II sources.  We
then first study the SED of selected models directly, and later turn
to a special color-color diagram in which we are able to display the
results of the full parameter study.

In Sect.~\ref{sec:model}, we describe the model
assumptions, take a first look at the resulting SEDs, and present the
basic model grid. In Sect.~\ref{sec:spectr-energy-distr}, we show the
SEDs for a selected subset of our parameter study.  In
Sect.~\ref{sec:color-color-diagram}, we describe the system to
quantify the SED structure using a color-color diagram, and
discuss in detail the parameter dependences of the model.  We discuss
some of the results in Sect.~\ref{sec:discussion} and present our
conclusions in Sect.~\ref{sec:conclusions}.

\section{Model setup}
\label{sec:model}

\subsection{The DD04 code}
\label{sec:dd04-code}
To model the proto-planetary disks of Herbig systems, we used the
code developed by \citet{d02} and \citet{dd04} in which 2-D radiative
transfer is solved subject to the constraint of radiative equilibrium
and vertical hydrostatic equilibrium for a passive disk (i.e. one in
which the only source of energy is irradiation by the central star)
for which the surface density structure is described by a power
law.

The code uses a spherical coordinate system $(r,\theta,\phi)$ with the
assumption of cylindrical symmetry (i.e. all derivatives with respect
to the azimuthal angel $\phi$ vanish).  The model consists of two
parts: a 2D radiative transfer code (RADICAL/RADMC) and a 1D vertical
hydrostatic equilibrium code. The radiative transfer code calculates
the dust temperature from the spatial distribution of the dust, and
the dust opacity.  The vertical hydrostatic equilibrium code calculates
the dust density from the dust temperature and vertical hydrostatic
equilibrium.  These two codes are applied alternately to an initial
guess for the dust density distribution until convergence is reached.
For details of the method we refer to the references given above.

No scattering is currently considered by the models.  Since at optical
wavelengths the scattering opacity for silicate grains (thought to be
a dominant dust species in proto-planetary disks) is approximately a
factor of a few larger than the absorption opacity (and is about equal
to it at near-IR wavelengths) one may expect that this may affect the
disk geometry and emerging spectral energy distribution.  \citet{dn03}
investigated this effect and concluded that because scattering
reflects part of the stellar light away from the disk before it
becomes thermalized, the disk becomes cooler at all vertical
heights. However, the vertical surface height -- defined to be the
surface at which the stellar light is intercepted -- will increase,
simply because of the added (scattering) opacity component. The effect
on the SED is that in the near- and mid-IR the flux may reduce by tens
of percents, while at millimeter wavelength the reduction is a few
percent at most. We conclude that neglecting the effects of scattering
may cause a modest (systematic) effect on the disk structure and
spectral appearance inferred, and therefore on the exact location of
the Group I/II boundary. However, it does not affect selectively
either Group I or II, and will therefore not affect any conclusions
regarding the cause of the Group I/II dichotomy itself.

\subsection{Modifications for the present study}
\label{sec:modif-pres-study}
The following modifications to the original code have been implemented
for the computations presented in this paper:

\subsubsection{Grain properties}
\label{sec:grain-properties}
In the pilot study by DD04, the grain population was assumed to consist
only of silicate particles with a radius $a = 0.1~\micron$.  Here, we
account for both silicate \citep{ld93} and carbon \citep{poy93} grains
that may range in size from a minimum value $a_{\rm min}$ to a maximum
value $a_{\rm max}$ according to
\begin{equation}
n(a) \propto \left( \frac{a}{a_{\rm min}} \right)^{-m}
\end{equation}
We adopt $m = 3.5$, which is representative of interstellar grains
\citep{mrn77}.  This power-law shape size distribution is expected on
theoretical grounds whenever grain-grain collisions may lead to
shattering \citep{1969JGR....74.2531D}, but differs from that expected
on the basis of grain growth
\citep[e.g.][]{2005A&A...434..971D,2005ApJ...625..414T} in which case
the dust size distribution has one or two peaks that shift with time.
If the distribution results from an equilibrium between growth and
fragmentation, it may also be flat \citep{2005A&A...434..971D}.  It is
therefore not at all obvious that a power law size distribution is
indeed the appropriate choice.  We used it nevertheless, because it is
the standard assumption in most disk models.  The relative abundance
of carbon and silicate grains is explored as a model parameter.

Grains of different sizes are all assumed to have the same
temperature.  Technically this is achieved by averaging the mass
absorption coefficients of the different grain sizes and materials,
weighted by the relative mass contained in each grain size and
material.  This averaged mass absorption coefficient is then used to
solve the radiative transfer problem.

\subsubsection{Small and large grains, and the \emph{small grain disk
    mass}}

Many studies of dust evolution in disks discuss the presence of
``large'' grains.  Depending on the type of study, this term may mean
grains of a few \um, as they may be detected by studying the 10\um{}
emission feature, or mm-sized grains as submm and mm studies report.
It is therefore important to define clearly what is meant by ``large''
and ``small'' in the context of a study.

In the present paper, we are interested mainly in the overall
structure of the disk model, and this structure will be dominated by
the grains that carry the opacity in the disk, i.e. the relatively
small grains.  Instead of using the total disk mass as an input
parameter, we therefore use the term \emph{small grain mass} of
the disk as the one main parameter of our disk models.  This  covers
all grains smaller than 25\um.  In most models we only consider
these grains and vary the small grain mass of the disk over a large
range.  This does not necessarily imply that the true dust mass in the
disk equals to this value.  In fact, one way to justify a large
range in the small grain mass is by assuming that much of the original
dust mass in the disk has been converted to larger grains, that
no longer influence the structure of the disk.  In
Sect.~\ref{sec:large-grains-mid}, we demonstrate a
computation of a few models in which we add significant amounts of dust
mass in the form of a midplane layer of large grains, and prove that
the general shape of the SED at near and mid-IR wavelength is largely
unaffected by this mass.

However, we study in detail the effects of changing the relative
abundance of grains below 25\um: these changes are expected to affect
both the strength of emission features close to 10\um, and the overall
disk structure.

\begin{figure}[t]
\centerline{\includegraphics[width=9cm]{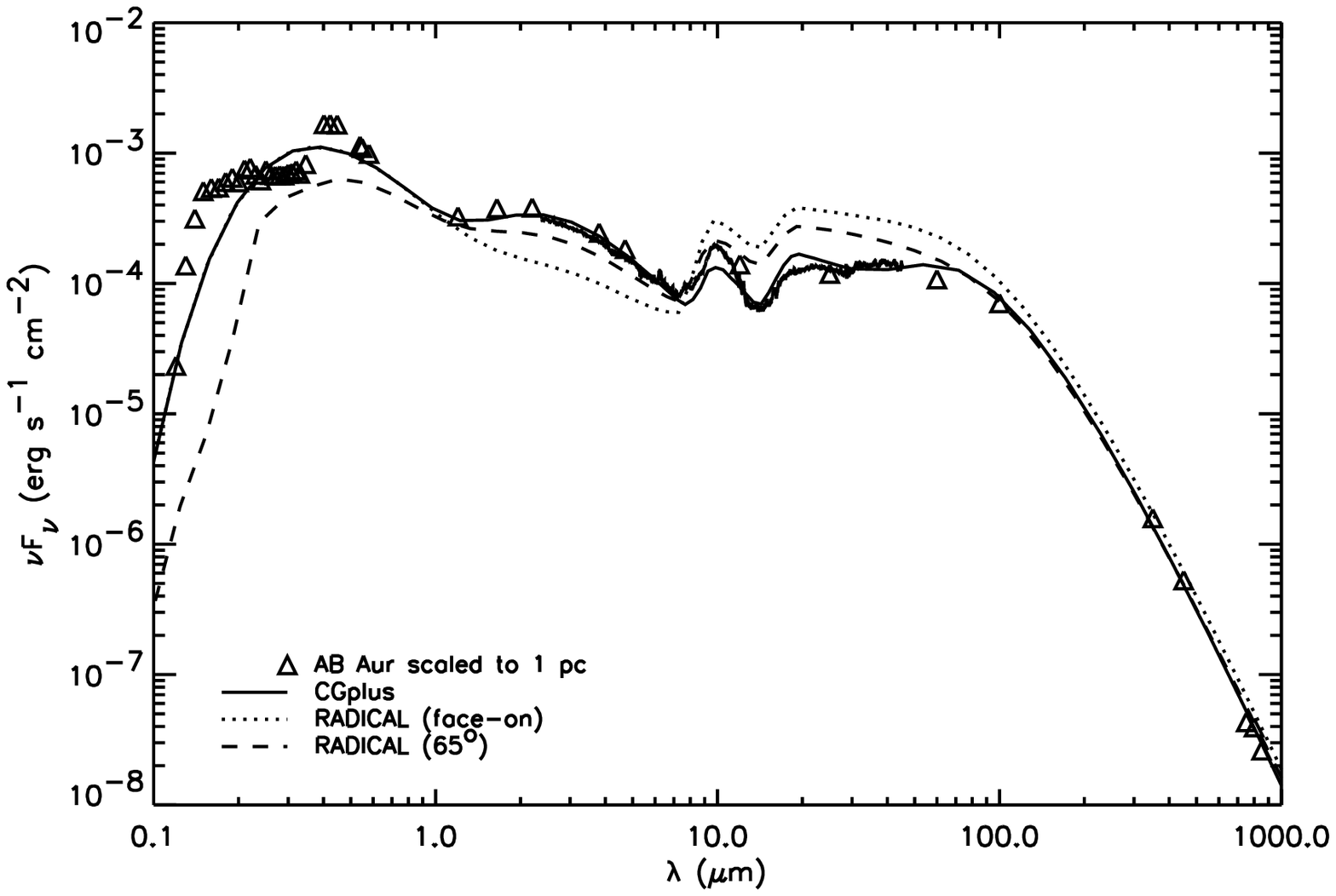}}
\caption[]{\label{fig:sed_abaur} The observed SED of the Herbig star
  AB Aurigae with several model fits.  The solid line shows
  the fit using the extended Chiang and Goldreich
  \citep{1999ApJ...519..279C} - like model including the effects of a
  puffed-up inner rim \citep{2001ApJ...560..957D}.  The fit was
  originally obtained by \citep{2003A&A...398..607D}, using an
  inclination of 65\degree.  The dotted line shows a face-on model
  obtained for the same star and disk parameters, the dashed line
  shows the same model tilted to a 65\degree inclination.}
\end{figure}

\subsection{First look: an example model and comparison to the DDN model}
\label{sec:first-look}
To obtain an impression of the new self-consistent models 
and how they compare with the DDN models, we use the model fit completed by
\citet[][hereafter D03]{2003A&A...398..607D} for the case of AB
Aurigae.  We have not made a specific fit with the new model to the
observed data.  For the sake of comparison, we have taken the
disk parameters exactly as they were derived by
D03, and solved the radiative transfer problem
and the vertical structure for this case.  In
Fig.~\ref{fig:sed_abaur}, we show the observed SED of AB Aurigae
and its D03 fit, the new model seen face-on, and the new
model seen at an angle of 65 degrees (the value derived by D03).

As can be seen, there are significant differences between the models,
most important of which may be the difference in the 3\um bump, which
represents the radiation that is caused mostly by the inner rim.
While by the D03 fit, this region of the SED is well fitted, the new
models underestimate the flux.  This is a systematic effect: the
self-consistent models under-predict this flux systematically, a fact
also noted by \citet{2006ApJ...636..348V}.  This is partly because the
temperature drops rapidly into the rim, and part of the absorbed
energy is emitted at wavelengths far longer than 3 \um.  It may be in
part because the geometric shape of the inner rim is in reality more
rounded than assumed here \citep{in05}.  This would also remove the
extreme dependence of the strength of the inner rim emission on the
inclination of the disk, as can be seen by comparing the model spectra
for a face on disk with one that is inclined by 65\degree.  The
unrealistically strong dependence on inclination is also the reason
why direct model fits infer high inclinations in out model - the inner
rim emission is highest at these angles.  \citet{in05} demonstrated
that a more rounded rim has its maximum flux in the face-on
orientation.  Unfortunately, we still need to devellop completely
self-consistent models for the inner rim.  We only need to notice that
there is a problem, and that inclination fits should not be trusted.
In Sect.~\ref{sec:inner-rim-systematics} we will consider the
systematic effects caused by the weak inner rim.

Another side-effect of the model at 65\degree inclination is that the
stellar flux is clearly reddened by the absorption of the outermost
disk parts.  At this inclination, the flaring outer part of the
disk extends to angles of up to 50\degree, so that the line-of-sight from
the observer to the star passes through these outer parts.  Dust
grains present there absorb the stellar light and redden the star
\citep[see, for example][]{2003ApJ...598.1079W}.

The far-infrared wavelength range of the SEDs emits more flux when
compared to the D03 model.  This is a direct consequence of the
different height of the inner rim.  In the D03 models, the higher rim
shadows the outer regions of the disk more, and consequently less
reprocessed radiation emerges from these regions.  Clearly, it would
be possible to construct a model that provides a better fit to the outer
regions, by reducing the small grain mass in the outer disk regions.
We have not attempted this here.

\begin{table}[tb]
   \caption{The parameters describing the disk structure and grain
     properties. The power law describing the grain size distribution
     has a fixed index $m = 3.5$, representative of interstellar
     grains. The values of the parameters that are not varied are
     marked by a $\star$.
   \label{tab:standardparameters}}
\setlength{\tabcolsep}{1.8mm}
\begin{tabular}{ll|rrrrr}
\multicolumn{2}{l|}{disk parameter}                        & \multicolumn{5}{c}{adopted values}                                             \\ \hline
\multicolumn{1}{l}{$\log$ dust mass [$M_{\odot}$]} & $M_{\rm dust}$  & -6.0                               & -5.5       & -5.0 & -4.5 & -4.0           \\
\multicolumn{1}{l}{outer radius [AU]}&$R_{\rm disk}$             & $^{*}$200                          & 400        & 600  & \multicolumn{2}{c}{ } \\
\multicolumn{1}{l}{mass distribution}& $p$                 & -1.0                               & $^{*}$-1.5 & -2.0 & \multicolumn{2}{c}{ } \\
\multicolumn{7}{l}{}\\
\multicolumn{2}{l|}{dust parameter}                        & \multicolumn{5}{c}{adopted values}                                             \\ \hline
size range [\micron]                                       & \multicolumn{1}{l|}{$a_{\rm min}$} & $^{*}$0.01 & 0.1  & 1    &                \\
                                                           & \multicolumn{1}{l|}{$a_{\rm max}$} & $^{*}$1    & 5    & 10   & 25             \\
\multicolumn{1}{l}{mass fraction}&$f_{\rm Sil}$           & $^{*}$1    & 0.95 & 0.91 & 0.83           \\
\multicolumn{1}{l}{}&$f_{Car}$                  & $^{*}$0    & 0.05 & 0.09 & 0.17           \\
\end{tabular}
\end{table}

\begin{figure*}[t]
\centerline{\includegraphics[]{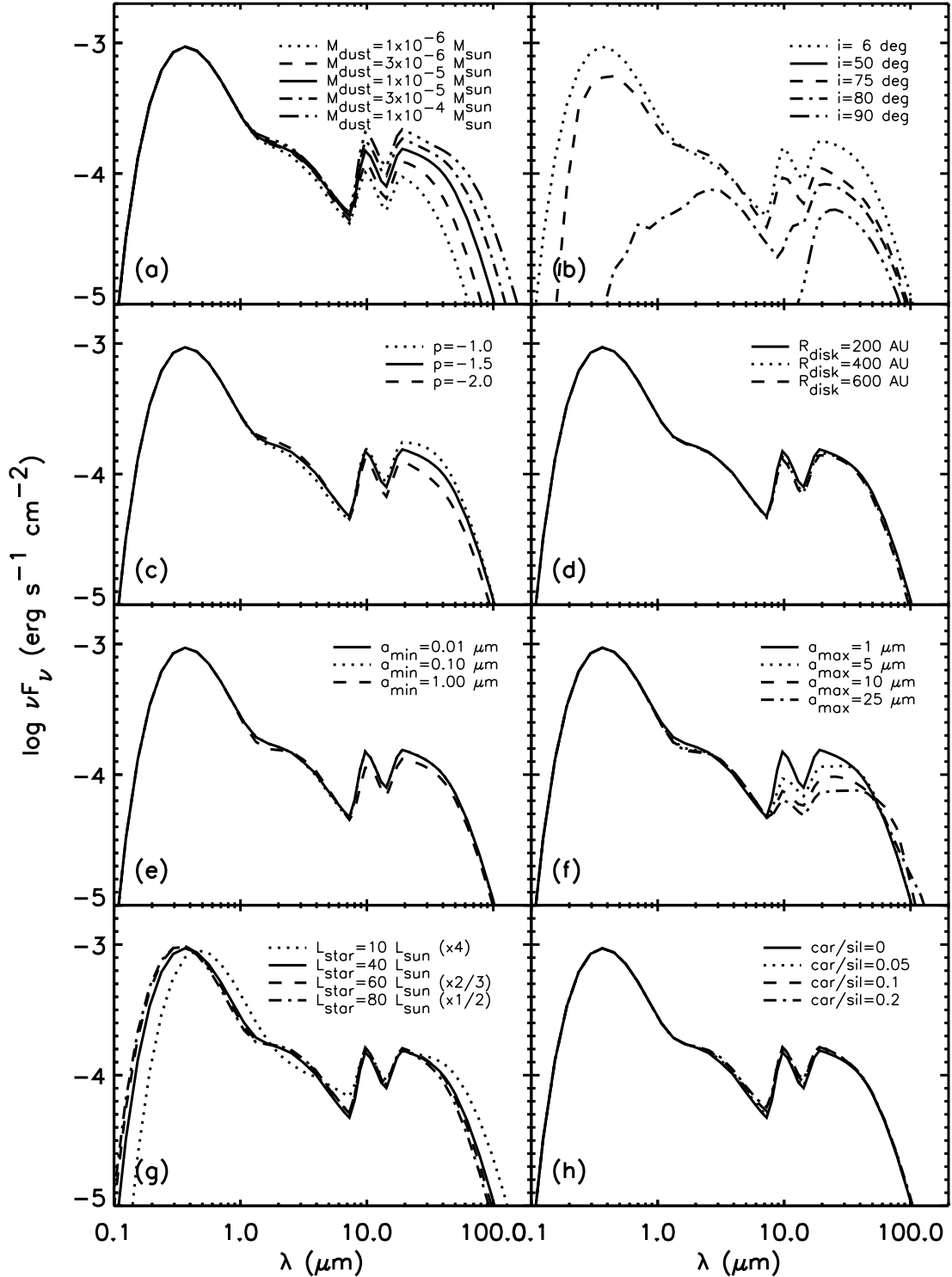}}
\caption[]{\label{fig:sedpanel} Panel representing an overview of the
  spectral energy distributions resulting from our parameter study.
  In each panel, we start from our standard base model
  (L$_{\star}=40L_{\odot}$, $M_{\rm dust}=10^{-5}M_{\odot}$,
  $i=50\degree$, $p=-1.5$, R$_{\rm disk}=200$AU, $a_{\rm
    min}=0.01\um$, $a_{\rm max}=1\um$) and vary a single parameter, as
  specified in the panels: (a) dust mass, (b) inclination, (c) index
  of power law of mass distribution, (d) disk size, (e) minimum grain
  size, (f) maximum grain size, (g) stellar luminosity, and (h)
  carbon/silicate ratio.}
\end{figure*}

\subsection{The model grid}
\label{sec:model-grid}


Our model grid was computed for a central star of mass $\mstar = 2.5
\,\msun$, radius $\rstar = 2.1 \,\rsun$ and effective temperature
$\teff = 10\,000$\,K, and therefore a luminosity $\lstar = 39.5
\,\lsun$. Its spectral energy distribution was represented by a Planck
function. An overview of the disk parameters is given in
Table~\ref{tab:standardparameters}. The free parameters describing the
global disk structure were: the total mass in small dust grains
\mdust, the mass distribution, and the outer disk radius \rdisk.  A
fixed gas-to-dust ratio $f = 100$ relates \mdust\ to the total disk
mass, although the amount of gas in the disk enters the calculation in
no way. The radial distribution of the surface density was given by
$\Sigma(r) \propto r^{-p}$, i.e. it is described by the power-law
index $p$ and the total disk mass.  We note that although the outer
disk radius is a free parameter, the inner disk radius is computed
self-consistently using the dust sublimation temperature \tsub. The
dust is assumed to consist of silicate and carbon grains. In the
current study, we were not interested in the effects of different
sublimation temperatures of different species.  We therefore assume
that both species sublimate together at \tsub\ = 1\,500 K, typical of
silicates.  We also investigated the effects of grain size and
computed a set of models with different grain size ranges (see
Table~\ref{tab:standardparameters}).  The full model grid therefore 
comprises 50 models. The relative abundance of the two dust species
considered was also varied, but the effect was negligible.

In Figs.~\ref{fig:sedpanel} and \ref{fig:panel}, we show the main
results of our parameter study.  Both figures contain 8 panels each,
and in each panel we study the dependence of the model spectra on one parameter.

\section{Spectral energy distributions}
\label{sec:spectr-energy-distr}
Figure~\ref{fig:sedpanel} shows the SEDs of the different models
relative to our standard model,which is given by the solid curve in each
panel.  This model uses a total mass in small dust grains of
10$^{-5}M_{\odot}$, a surface density power law with $p=-1.5$, a
dust size distribution between 0.01 and 1.0\um, a carbon/silicate mass
ratio of 0 and a luminosity of 40L$_{\odot}$.  The SED of
the standard model is computed at an inclination of 50\degree.
In each panel, we then vary one parameter, as specified by the labels
in the diagram.

\subsection{Overview over SED parameter dependence}

Figure~\ref{fig:sedpanel} shows the spectral energy distributions for
a subset of our parameter study.  In each panel, we start from our
standard base model and then vary a single parameter.  The resulting
SEDs are shown, and the different values of the varied parameter is
shown in the panel legend.  We present a far more detailed
discussion of the parameter dependencies in
Sect.~\ref{sec:gener-param-trends};  here, we only highlight
the obvious properties of these calculations.  It is
clear that the dependence of the SED on the mass of small
dust grains (panel a), the inclination (panel b) and the maximum grain
size (panel f) is far stronger than the influence of all other
parameters.  In panels d, e, and h, hardly any variations are seen,
whereas the data in panels c and g show moderate changes.

One result is particularly significant:  The shape and strength
of the 3\um bump is extremely robust and similar in all models, with
only two exceptions.  One exception is high inclination models
($i>75$\degree).  In these models, the inner rim is seen through the
outer disk, and the 3\um flux is strongly absorbed.  The other exception
is the model with a luminosity of only 10L$_{\odot}$ for which the
inner rim emission is less significant.  The reason for this
latter case is that at low luminosities, the dust evaporation radius
is far closer to the star, and therefore deeper inside the gravitational
potential of the star.  The gravity of the star reduces the ratio of
surface height to distance, so that a smaller fraction of the total
luminosity is absorbed by the inner rim.  Apart from those exceptions,
this part of the spectrum shows an amazing lack of variation.  This is a
reflection of two important facts: (i) the distance of the rim is set
by evaporation physics and therefore independent of the disk
properties. (ii) its height is only weakly dependent on disk
properties because the rim is \emph{strongly} optically thick and
remains so regardless of parameter changes.

The SED variations in panel b (inclination) only
occur because of the location of the observer relative to the object.
If we focus on the remaining parameters that alter the intrinsic
properties of the star-disk system, the strongest influence is exerted
by the mass in small grains, which can switch the outer disk from
optically thick to optically thin.

Some of the variations in panel f are due to a significant reduction
in the optical depth of the outer disk, which is caused by an increase
in the upper limit of the dust grain size distribution to 25\um.  In
addition, the 10\um feature is also weakened because this feature is mostly
due to grains with sizes below 3\um.

\begin{table}[tb]
   \caption{Dust mass in small ($0.01-1{\rm~\mu m}$) and large ($2
            {\rm~mm}$) grains for the mid-plane layer grid. The star
            has the following parameters: $M_*=2 {\rm~M_{\sun}}$,
            $R_*=3 {\rm~R_{\sun}}$ and $T_*=10\,000 {\rm~K}$, and thus
            $L_*=80.7 {\rm~L_{\sun}}$. The disk has a size of $200
            {\rm~AU}$ and a dust mass of $5\times10^{-4}
            {\rm~M_{\sun}}$. The small dust in the disk has a grain
            size range of $0.01-1{\rm~\mu m}$ with a power-law
            distribution with a power of $-3.5$. The large dust in the
            mid-plane layer has a size of $2 {\rm~mm}$.
   \label{tab:layerparameters}}
\begin{tabular}{lccc}
      & $M_{\rm dust, orig}$ & $M_{\rm mid plane}$ & large grain   \\
model & [$M_{\odot}$]        & [$M_{\odot}$]         & mass fraction \\\hline
ML0   & $1\times10^{-6}$     & $4.99\times10^{-4}$ & $0.998$       \\
ML1   & $3\times10^{-6}$     & $4.97\times10^{-4}$ & $0.994$       \\
ML2   & $1\times10^{-5}$     & $4.90\times10^{-4}$ & $0.980$       \\ 
ML3   & $3\times10^{-5}$     & $4.70\times10^{-4}$ & $0.940$       \\
ML4   & $1\times10^{-4}$     & $4.00\times10^{-4}$ & $0.800$       \\
ML5   & $5\times10^{-4}$     & $0$                 & $0.000$       \\
\end{tabular}
\end{table}

\begin{figure}[t!]
\centerline{\includegraphics[width=8cm]{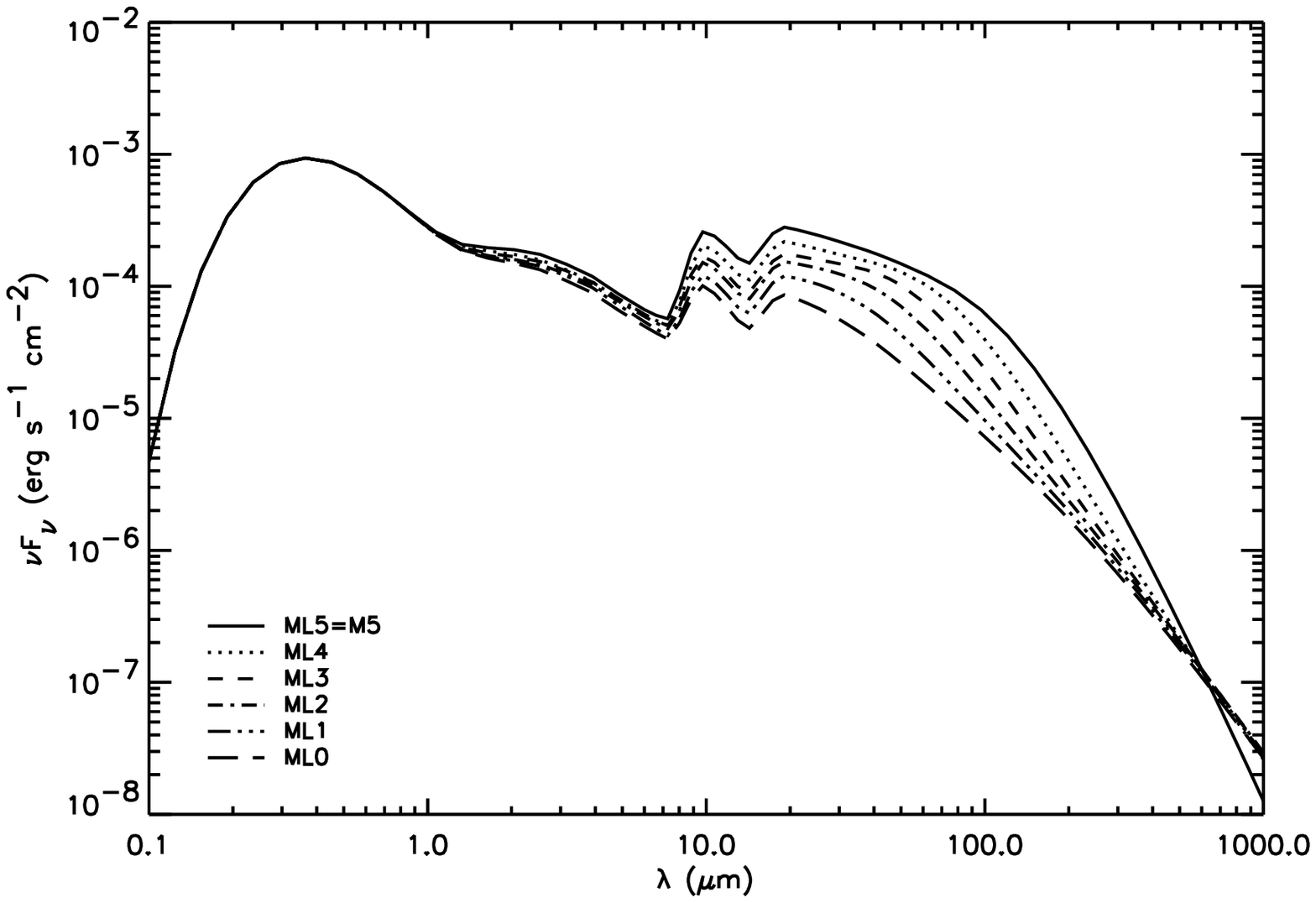}}
\centerline{\includegraphics[width=8cm]{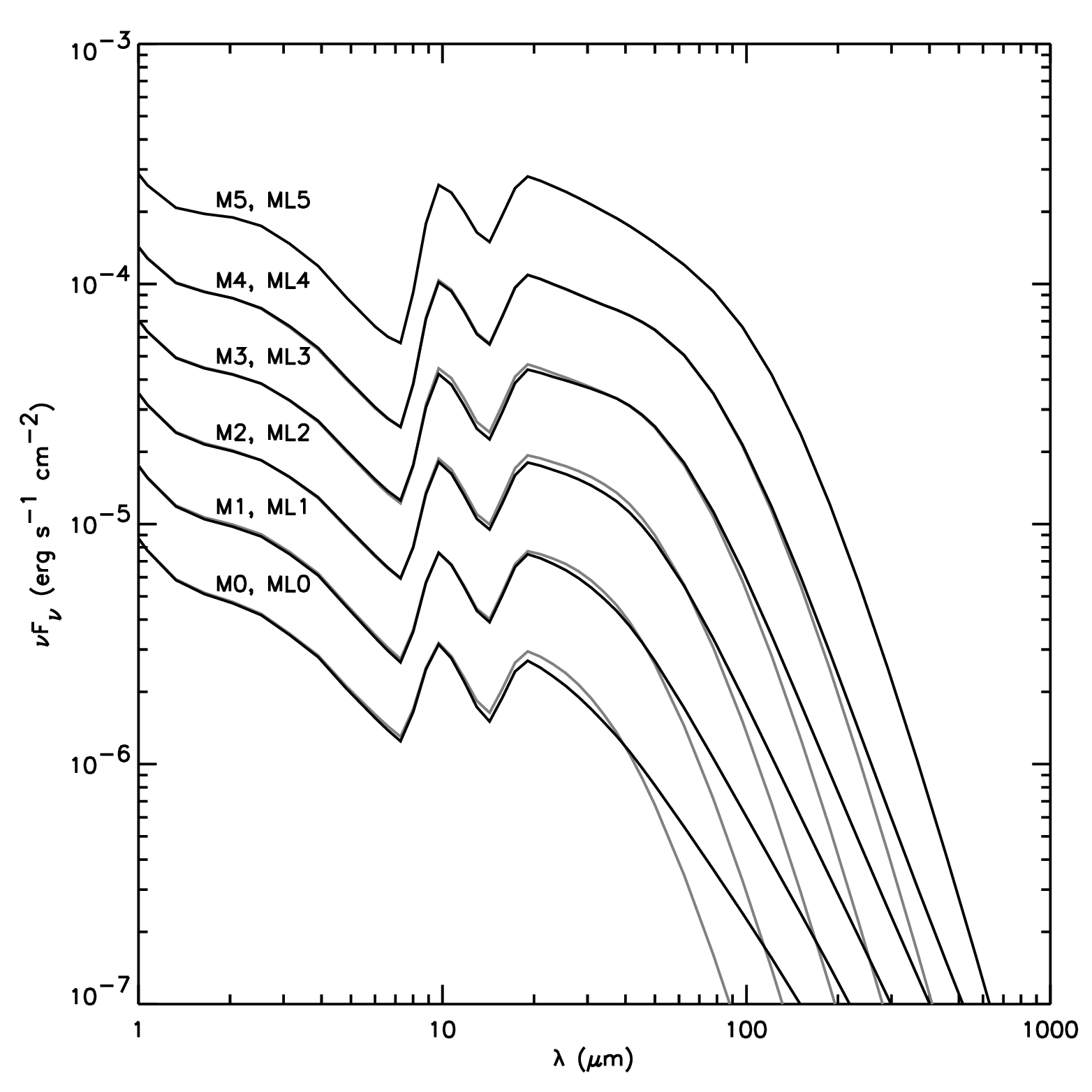}}
\caption[]{\label{fig:sed-layers} The SEDs of the models in which a
  midplane layer of large grains is added so that the total dust mass
  in the disk is kept constant.  The upper panel shows the models
  ML0\ldots ML5 that display a transition from group I to group II SED
type, but all converge again at about 800\,\um.  The lower panel
compare each pair of models with equal small grain mass.  The SED's
are shifted vertically for improved visibility.}
\end{figure}

\subsection{Large grains in the mid plane}
\label{sec:large-grains-mid}

We considered a wide range of total small dust grain mass in the disk
in order to produce models of significantly different properties.  In
reality, the amount of solids in a disk might span a far smaller range
ov values \citep{aa04}, and the small dust mass required to reproduce
group II models may be compensated for by larger grains or even
pebbles and boulders in the disk midplane.  DD04 already demonstrated
that converting a large fraction of small grains into large midplane
particles produces models that are similar to those of small dust
mass, apart from the fact that the submm luminosity is increased.
Therefore, throughout this paper, low dust mass can be assumed to
correspond to models with an efficient conversion of micron-sized dust
to far larger particles.

To assess the effect of sedimentation on the spectral energy
distribution, we calculated an additional five models, in which we introduced a
mid-plane layer of large grains. This was achieved by taking a model from
the standard grid with a dust mass less than $5\times10^{-4}
{\rm~M_{\sun}}$ and adding a mid-plane layer of $2 {\rm~mm}$ grains
of a mass such that the total disk mass equaled $5\times10^{-4}
{\rm~M_{\sun}}$. The resulting values of the mass in the midplane
layer can be seen in Table~\ref{tab:layerparameters}.  The large grain
mass fraction varies between $0.998$ for model ML0 and $0.000$ for
model ML5, which is equivalent to its parent model M5 (see
Table~\ref{tab:models}).

The SEDs are shown in Fig.~\ref{fig:sed-layers}.  In the upper
panel, we compare all models that include a mid-plane layer.  While
there are strong differences in the near, mid, and far infrared, the
models converge  in the submm region, indicating equal total
mass.  In the lower panel, we compare two models with the same
mass in small grains, one without the added midplane layer, one with
the added layer.  It is clear from these plots that the shape
of the SED blueward of about 60\,\um is determinated by the mass
in small grains, while at wavelengths longer than 100\,\um clear
differences can be seen.  In a classification of SEDs based on the
SED up to 60\um, we can therefore ignore the presence or absence of a
mass-compensating midplane layer of large grains.  We also demonstrate
this in Sect.~\ref{sec:infl-large-midpl}.

This implies that sources with the same small grain mass have the same
group classification, but not necessarily the same total dust mass
because mass may be 'hidden' in large grains.  Since large grains
affect the shape only at sub-mm wavelengths and are therefore not
evident in the \citet{bwd03} diagram, adding large grains in the mid
plane of the disk does not influence the Meeus classification
significantly.  We can therefore conclude that differences between the
properties fo group I and a group II sources are due to the small ($a
< 1~\micron$) grain mass.  While the models infer a factor of 100
difference in sub-mm flux between group I and II, observations only
show a factor of 10. This implies that large grains must exist in the
group II sources to increase the sub-mm flux to the observed
levels. As demonstrated by \citet{aad04}, the sub-mm slope of
group II sources also indicates that there must be more large grains
present than in group I sources.

\section{Color-color diagram}
\label{sec:color-color-diagram}
\subsection{Quantifying SED classification}
\label{subsec:quantify}

To quantify the amount of flaring in all calculated models, we
needed to employ a method that is both fast and accurate.  Previously, the
classification of group I/II was achieved by fitting power-law and/or
black-body functions to the SED \citep{mwb01}, splitting the SEDs
into two groups based on the presence (group I) or absence (group II)
of a black-body component in addition to the power-law component.  Since
this proved to be an inefficient method when analyzing hundreds of SEDs, we
chose a simple quantitative way of classifying SEDs.  BWD03 provided
us with such a method.

\begin{figure}[t]
\centering
  \includegraphics[width=\columnwidth]{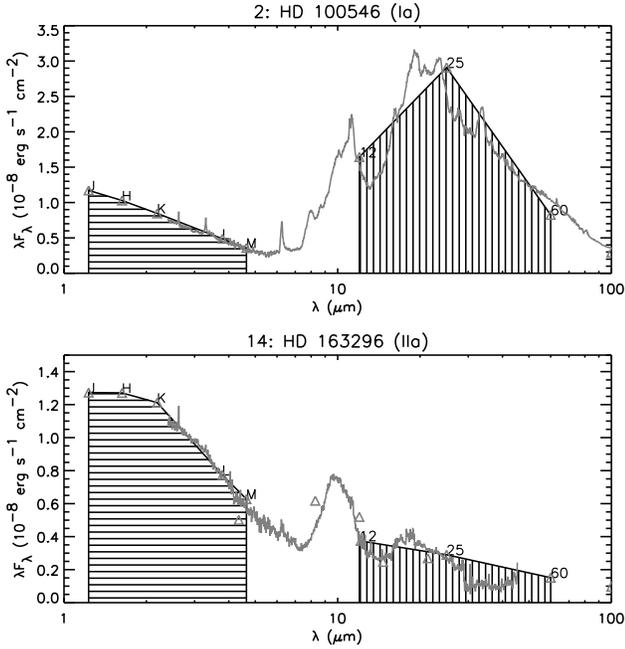}
   \caption{SEDs of two sources HD 100546 and HD 163296 showing the
     integrated fluxes $L_{\rm NIR}$ and $L_{\rm FIR}$. The ratio of
     these integrated fluxes and the IRAS $m_{12}-m_{60}$ color is
     used in all van Boekel diagrams in this paper.}
      \label{fig:sed_onetwo}%
\end{figure}

\begin{figure}[t]
\centering
  \includegraphics[width=\columnwidth]{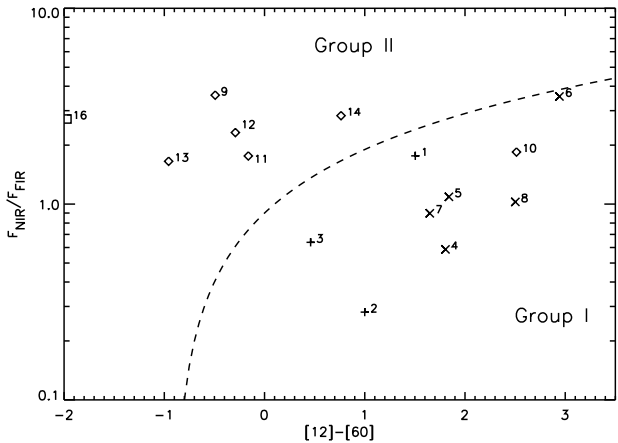}
  \caption{The \citet{bwd03} diagram for the sources listed in
    Table~\ref{tab:observations}. On the ordinate, the IRAS
    $m_{12}-m_{60}$ color and, on the abscissa, the $L_{\rm NIR}/L_{\rm
      FIR}$ flux ratio is given. The numbers of the observations
    correspond to the numbers in Table~\ref{tab:observations}.  Pluses
    indicate Ia sources, crosses Ib sources, diamonds IIa sources, and
    the square a group IIb source.  We note that all Ib sources are in
    the upper right corner of the group I zone. The dashed line
    represents $L_{\rm NIR}/L_{\rm FIR} = (m_{12}-m_{60}) + 0.9$.}
      \label{fig:onetwo}%
\end{figure}

In BWD03, an IRAS $m_{12}-m_{60}$ color versus $L_{\rm NIR}/L_{\rm FIR}$
diagram was presented.  $L_{\rm NIR}$ was the integrated luminosity from
JHKLM photometry and $L_{\rm FIR}$ the same quantity derived from the
IRAS 12, 25, and 60~\micron\ data points.  Figure~\ref{fig:sed_onetwo}
shows these two luminosities for two sources typical of group I and
group II. The IRAS $m_{12}-m_{60}$ color is sensitive to the slope of
the SED between 12 and 60 $\mu$m, which is steeper in group II sources
resulting in a lower IRAS $m_{12}-m_{60}$ color for group II sources
than group I sources. The near to far infrared luminosity ratio
$L_{\rm NIR}/L_{\rm FIR}$ is sensitive to the disk surface-temperature
distribution. Since group I sources have warmer outer disk surface
temperatures than group II sources their FIR emission will be higher
relative to the NIR emission and the NIR over FIR ratio will be
lower.  Data for group II sources occupy mainly the upper left corner
whereas group I sources will occupy primarily the lower right corner. An
example of the diagram is shown in Fig.~\ref{fig:onetwo}.  The
sources used in Fig.~\ref{fig:onetwo} are listed in
Table~\ref{tab:observations}.

The line $L_{\rm NIR}/L_{\rm FIR} = (m_{12}-m_{60}) + 0.9$ was
determined to be the most reliable means of separating group I from
group II sources. Sources above the line belong to group II, sources
below the line to group I.  HD 142527 (\#11 in
table~\ref{tab:observations}) does not comply with this rule because
of its different geometry
\citep{2006ApJ...636L.153F,2006ApJ...644L.133F}.  However, the
porperties of inner part of its disk indicates that it is group II
\citep{2004A&A...423..537L}.

%

\begin{table*}[tb]
   \caption{A list of sources that were used to compare with our
     models. In Col.\ 1 we give the identification number that appears in all 
            diagrams. Column 2 contains the name of the star. In
            Col.\ 3, 4, 5, and 6, the distance, effective temperature,
            stellar luminosity and stellar mass are given. This data
            was taken from \citet{bmw05}. In Col.\ 7, the
            stellar radius is given. In Col.\ 8, we indicate the IRAS
            $m_{12}-m_{60}$ color. Column 9 lists the flux-ratio of NIR over
            FIR. The NIR flux is determined from JHKLM photometry. The
            FIR flux is derived from IRAS 12, 25 and 60 fluxes. In
            Col.\  10 and 11 the ratios of NIR and FIR flux to stellar
            flux are given.  In Col.\ 12 the group classification
            according to \citet{mwb01} is given.
            \label{tab:observations}}
\begin{tabular}{rlrrrrrrllll}
(1)&    (2)    &  (3)  &   (4)   &   (5)   &  (6)   &  (7)  &   (8)    &   (9)   &  (10)   &   (11)  & (12) \\
\# & star & $d$        & $T_{\rm eff}$ & $L$              & $M$              & $R$              & $[12]-[60]$ 
& $F_{\rm NIR}/F_{\rm FIR}$ & $F_{\rm NIR}/F_{\star}$ & $F_{\rm FIR}/F_{\star}$ & group \\ 
   &      & $\rm [pc]$ & $\rm [K]$     & $\rm [L_{\sun}]$ & $\rm [M_{\sun}]$ & $\rm [R_{\sun}]$ & & & & & \\\hline\hline
 1 & AB Aur    & $144$ &  $9528$ &  $47.0$ & $2.4$  & $2.5$ &  $1.504$ & $1.763$ & $0.274$ & $0.155$ & Ia  \\
 2 & HD 100546 & $103$ & $10495$ &  $32.0$ & $2.4$  & $1.7$ &  $1.000$ & $0.281$ & $0.113$ & $0.404$ & Ia  \\
 3 & HD 179218 & $243$ & $10495$ & $100.0$ & $2.9$  & $3.0$ &  $0.460$ & $0.638$ & $0.099$ & $0.155$ & Ia  \\\hline
 4 & HD 97048  & $175$ & $10000$ &  $43.7$ & $2.5$  & $2.2$ &  $1.806$ & $0.587$ & $0.095$ & $0.162$ & Ib  \\
 5 & HD 100453 & $111$ &  $7396$ &   $7.9$ & $1.7$  & $1.7$ &  $1.840$ & $1.089$ & $0.287$ & $0.263$ & Ib  \\
 6 & HD 135344 & $140$ &  $6592$ &   $8.1$ & $1.6$  & $2.2$ &  $2.943$ & $3.545$ & $0.360$ & $0.102$ & Ib  \\
 7 & HD 139614 & $140$ &  $7852$ &   $8.1$ & $1.7$  & $1.5$ &  $1.649$ & $0.897$ & $0.191$ & $0.213$ & Ib  \\
 8 & HD 169142 & $145$ &  $8204$ &  $14.5$ & $2.0$  & $1.9$ &  $2.503$ & $1.026$ & $0.136$ & $0.133$ & Ib  \\\hline
 9 & HD 104237 & $116$ &  $8414$ &  $35.0$ & $2.3$  & $2.8$ & $-0.492$ & $3.599$ & $0.252$ & $0.070$ & IIa \\
10 & HD 142527 & $198$ &  $6252$ &  $29.0$ & $2.5$  & $4.6$ &  $2.513$ & $1.844$ & $0.417$ & $0.226$ & IIa \\
11 & HD 142666 & $145$ &  $7580$ &  $13.5$ & $1.8$  & $2.1$ & $-0.162$ & $1.761$ & $0.222$ & $0.126$ & IIa \\
12 & HD 144432 & $145$ &  $7345$ &  $10.2$ & $1.8$  & $2.0$ & $-0.291$ & $2.319$ & $0.324$ & $0.140$ & IIa \\
13 & HD 150193 & $150$ &  $8974$ &  $24.0$ & $2.3$  & $2.0$ & $-0.957$ & $1.654$ & $0.212$ & $0.128$ & IIa \\
14 & HD 163296 & $122$ &  $8730$ &  $24.0$ & $2.0$  & $2.2$ &  $0.763$ & $2.827$ & $0.286$ & $0.101$ & IIa \\
15 & 51 Oph    & $131$ & $10000$ & $245.0$ & $2.0$  & $5.1$ & $-2.924$ & $9.294$ & $0.066$ & $0.007$ & IIa \\\hline
16 &  HD 95881 & $118$ &  $8990$ &  $11.6$ & $1.7$  & $1.4$ & $-1.971$ & $2.719$ & $0.203$ & $0.074$ & IIb \\
\end{tabular}
\end{table*}

\begin{table*}[tb]
\caption{Flux ratios and infrared colors for selected models.  The
  first sections contains the different AB Aur models shown in
  Fig.\ref{fig:sed_abaur}.
The second section shows the values for the standard models with
different small grain masses.  The final section shows values derived
from the models with a midplane layer of large grains.
  \label{tab:models}}
\begin{tabular}{lrrrrrrrrrrrr}
model & $M_{\rm dust}$ & $i$ & [12]-[60] & $F_{\rm NIR}/F_{\rm FIR}$ & 
$F_{\rm NIR}/F_{\rm *}$ & $F_{\rm FIR}/F_{\rm *}$\\
 & $M_{\odot}$ &  &  &  &  & \\\hline\hline
 CGplus & $5(-2)$  & 65 &  2.158 &  1.680 & 0.273 & 0.162\\
RADICAL & $1(-3)$  &  6 &  1.793 &  0.401 & 0.142 & 0.355\\
RADICAL & $1(-3)$  & 65 &  1.671 &  0.802 & 0.206 & 0.256\\\hline
     M0 & $1(-6)$  & 50 &  0.08  &  1.60  & 0.156 & 0.097 \\
     M1 & $3(-6)$  & 50 &  0.61  &  1.26  & 0.164 & 0.130 \\
     M2 & $1(-5)$  & 50 &  1.01  &  1.00  & 0.167 & 0.167 \\
     M3 & $3(-5)$  & 50 &  1.24  &  0.86  & 0.173 & 0.202 \\
     M4 & $1(-4)$  & 50 &  1.41  &  0.74  & 0.177 & 0.239 \\
     M5 & $5(-4)$  & 50 &  1.34  &  0.63  & 0.190 & 0.303 \\\hline
    ML0 & $5(-4)$  & 50 &  0.42  &  1.69  & 0.154 & 0.091 \\
    ML1 & $5(-4)$  & 50 &  0.72  &  1.27  & 0.161 & 0.127 \\
    ML2 & $5(-4)$  & 50 &  1.01  &  0.99  & 0.165 & 0.166 \\
    ML3 & $5(-4)$  & 50 &  1.31  &  0.89  & 0.173 & 0.194 \\
    ML4 & $5(-4)$  & 50 &  1.42  &  0.74  & 0.177 & 0.238 \\
\end{tabular}
\end{table*}

Other aspects of the disk and the dust also play a role in determining
the position of a model in this diagram. A model without a puffed-up
inner rim emits less NIR and more FIR flux than a model with a
puffed-up inner rim. This is because the puffed up inner rim
re-emits more radiation in the NIR and casts a shadow on the
disk thereby decreasing the FIR flux. The FIR flux will decrease more
on the blue side and increase in slope because the shadow does not affect
the outer parts of the disk in which the flux redward of 20~\micron\
originates.  Therefore, a model without a puffed-up inner rim has
a lower NIR/FIR ratio and a redder IRAS $m_{12}-m_{60}$ color
and will be shifted down and to the right in the diagram with respect to
the models with a puffed-up inner rim.

Adding a mid-plane layer of large grains to a model will increase the
FIR flux longward of approximately 50~\micron. The large grains in the
mid plane are only visible in the outer parts of the disk, where they
are cold and emit at longer wavelengths.  This means that adding a
mid-plane layer will increase both the FIR flux and the IRAS
$m_{12}-m_{60}$ color, shifting the predicted data points for a model
with an added mid-plane layer of large grains to the right of and
slightly lower than the original model.

The NIR and FIR windows were chosen so that there is only a
limited contribution from the silicate emission around 10~\micron\ and
it is primarily the geometry of the disk determine the position
in the diagram. However, (cold) crystalline silicates also emit a
broader and less conspicuous feature that peaks around 20~\micron. If
there is a large amount of crystalline silicates in the outer part of
the disk (beyond 2 AU), such as that in HD 142527
\citep[see][]{bml04}, this will contribute significantly to the FIR
flux. This means that sources with highly crystalline (outer) disks
will appear lower in the diagram.

\begin{figure*}[t!]
\centering
  \includegraphics[width=0.75\textwidth]{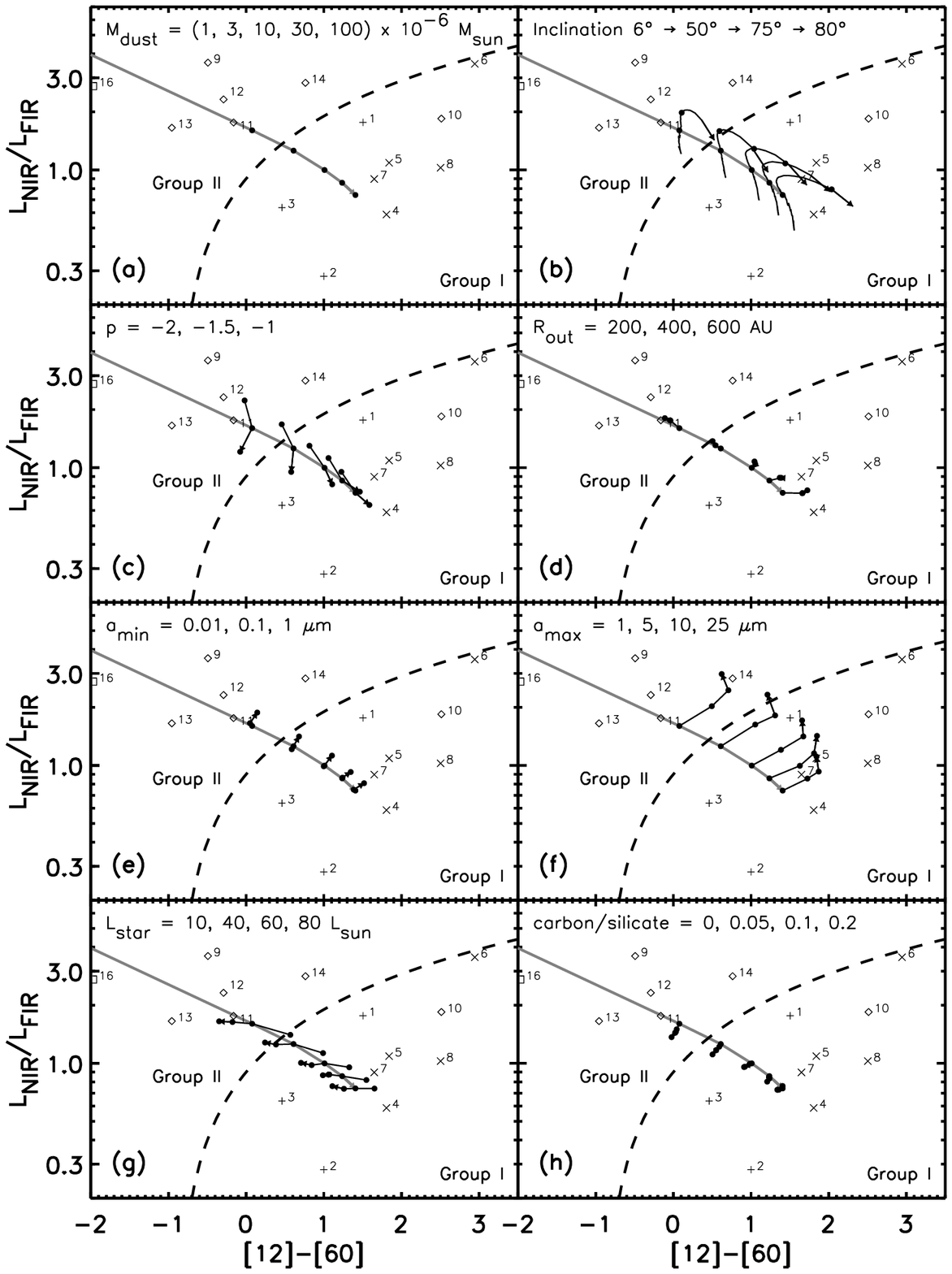}
  \caption{Plot with \citet{bwd03} diagrams, for variations of the
    most important parameters.  In each panel, the dashed line
    represents $L_{\rm NIR}/L_{\rm FIR} = (m_{12}-m_{60}) + 0.9$, the
    line used empirically to classify sources.  The standard models
    which are shown in the first panel have an inclination of
    50\degree, a surface density distribution power law of $p=-1.5$,
    an outer radius of 200 AU and a grain-size distribution ranging
    from 0.01 to 1~\micron.  The parameter varied in each panel is:
    {\em a) dust mass} Dust mass increases to the right. The grey line
    connects the five standard models and one optically thin model
    that lies outside the diagram.  This line is present in all
    panels. {\em b) Inclination} The standard models are at the
    average inclination of 50\degree.  Inclination increases along the
    curve from the bottom to the right. The triangle in each curve
    indicates an A$_{\rm V}$ of 0.1 mag, the point when the outer disk
    starts to significantly absorb the radiation coming from the inner
    rim. {\em c) index of power law of mass distribution} The power
    law becomes shallower from top to bottom.  {\em d) disk size}.
    {\em e) minimum grain size}.  {\em f) maximum grain size}.  {\em
      g) Stellar luminosity}. {\em h) Carbon-Silicate ratio.}}
      \label{fig:panel}%
\end{figure*}

Figure~\ref{fig:panel} shows the position of all models in our
parameter study in the diagnostic color-color diagram.  We varied
each parameter for 6 different small-grain dust masses, using the
models M0--M5 as the basis for each variation track.

%

\subsection{Influence of large midplane layer grains}
\label{sec:infl-large-midpl}

\begin{figure}[t]
\centering
  \includegraphics[width=\columnwidth]{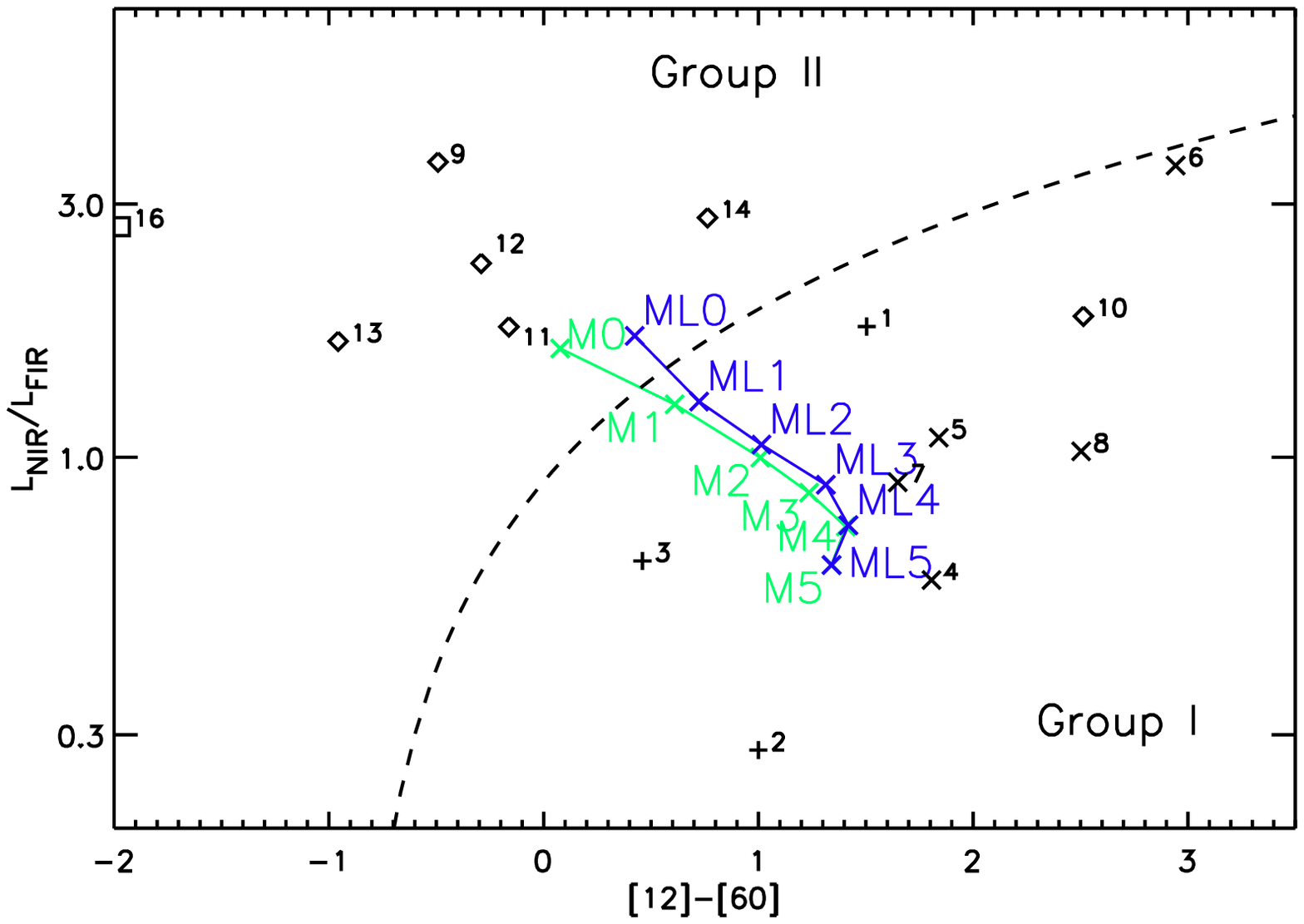}
   \caption{IRAS $m_{12}-m_{60}$ colour versus $L_{\rm NIR}/L_{\rm
            FIR}$ diagram with two different grids of models. Models ML0
            to ML5 contain a mid-plane layer of large grains.  The
            properties of this layer can be seen in
            table~\ref{tab:layerparameters}. Models M0--M5 are the
            original models, to which the mid-plane layer was added.
      \label{fig:layer}}
\end{figure}

In Fig.~\ref{fig:layer} we show the positions in the diagnostic
diagram of the models M0--M5, and compare them with the positions of
the corresponding models in which a low value of the small grain mass
is compensated by a layer of large grains in the mid-plane.  This
figure clearly confirms our expectation from
Sect~\ref{sec:large-grains-mid}.  The positions of the models with the
same small-grain mass are very close to each other.  Models ML2--ML5
occupy exactly the same position as the corresponding models M2--M5.
A visible shift occurs only for models with the lowest small grain
masses, ML0/M0 and ML1/M1.  However, the shift is small compared with
the overall range for both observations and modelling results in this
diagram.  We may therefore restrict our discussion to the small
grain masses.

\subsection{General parameter trends}
\label{sec:gener-param-trends}
We will discuss the trends observable in each panel of
Fig~\ref{fig:panel} and the physics in the models responsible for
these trends.
\begin{description}


\item[\bf Dust mass] In Figs.~\ref{fig:sedpanel}(a) and
  \ref{fig:panel}(a), the effect on the spectrum of increasing the
  dust mass can be seen. As concluded before in DD04, dust mass is
  an important factor in determining the disk geometry.  Higher mass
  provides hgher opacity, which causes more flaring.  An increasing
  dust mass therefore shifts a model from group II to group I. It should be
  noted, however, that primarily small grains ($<$10~\micron) contribute
  to the opacity in the disk, so the dust mass mentioned is
  the mass in small dust grains.  This is more evident when we
  add a mid-plane layer of large (2~mm) grains in
  Fig.~\ref{fig:layer}.

\item[\bf Inclination] In Figs.~\ref{fig:sedpanel}(b) and
  \ref{fig:panel}(b), the effect of changing the inclination on the
  spectrum of models with different dust mass is shown. Increasing the
  inclination from face-on to the angle at which A$_{\rm V}$ equals
  0.1 (indicated by a triangle) enables more of the vertical inner rim
  to become visible. This causes the NIR flux to increase.
  Furthermore, increasing the inclination caused the outer part of the
  disk to absorb an increasing amount of the inner rim radiation.
  This decreases the NIR flux, and reddens the entire
  spectrum making the IRAS $m_{12}-m_{60}$ color increase.  At 
  extreme inclination angles ($>80\degree$, not shown here), the
  reddening becomes so severe that the SED would no longer be
  classified as a (Lada-)class \citep{Lada87} II source, but rather as
  a class I source.

\item[\bf Power-law of Mass Distribution] In
  Figs.~\ref{fig:sedpanel}(c) and \ref{fig:panel}(c), we can observe
  the effect of changing the surface density power law.  Steepening
  the power law while keeping the disk mass the same, causes
  relatively more of the mass to be on the inside of the disk. Between
  the shallowest ($p=-1.0$) and the steepest ($p=-2.0$) power law
  slope, the mass in the inner 1 AU increases by a factor of 46 and
  the height of the inner rim by a factor of two. Having relatively
  more mass on the inside implies that there is less mass on the
  outside where the FIR flux originates.  Thus the FIR flux decreases
  while the NIR flux increases and the NIR/FIR ratio increases as the
  power-law slope steepens.  In the higher mass models the mass
  in the outer disk region is colder and emits mainly redward of
  20~\micron.  The decrease in FIR flux blue-shifts the IRAS
  $m_{12}-m_{60}$ color value.

\item[\bf Outer Radius]In Figs.~\ref{fig:sedpanel}(d) and
  \ref{fig:panel}(d), the effect of an increasing outer disk radius is
  shown. Increasing the disk size while keeping the dust mass the same
  causes more of the mass to be on the outside and less on the
  inside. However, the mass is redistributed far less significantly
  than when the power-law slow is altered. The mass in the inner AU reduces by
  only a factor of two and the mass on the outside shows no
  significant increase. To achieve a similar redistribution of mass as
  in the power-law models, we would have to increase the disk size to
  $4\times10^{5}$~AU.  The position of the models in the diagram would then
  hardly change.

\item[\bf Minimum grain size] In Figs.~\ref{fig:sedpanel}(e) and
  \ref{fig:panel}(e), the effect of increasing the lower limit on
  of the grain-size distribution is shown. Because of the $m=-3.5$
  power-law slope in the size distribution we use, most mass is in the
  largest grains.  Increasing the minimum grain size therefore has
  little effect on the SED.  The small grains emit mostly in the
  10-20~\micron\ wavelength range. This affects the FIR flux and the
  12~\micron\ IRAS color, and the NIR/FIR and the IRAS
  $m_{12}-m_{60}$ color therefore increase.

\item[\bf Maximum grain size] In Figs.~\ref{fig:sedpanel}(f) and
  \ref{fig:panel}(f), the effect of increasing the upper limit on the
  grain-size distribution is shown.  By increasing the upper boundary
  of the size distribution power law while keeping the lower boundary
  constant, mass is extracted from the population of smaller grains
  and placed into the larger ones.  Large grains have lower opacity
  (absorption cross section \emph{per mass}).  Consequently, the
  overall IR flux decreases and the 10\micron\ and 20\micron\ features
  reduce in strength.  The FIR flux decreases more rapidly than the
  NIR flux because the outer parts of the disk become optically thin
  (leading to weaker flaring), while the inner parts hardly change in
  structure.  Beyond 60~\micron, the flux increases as more and more
  cold material is revealed.  However, when the maximum grain size
  exceeds 10~\micron, the flux also decreases in this part of the SED
  because the decrease in opacity becomes significant.  This effect
  causes the reversal in the IRAS $m_{12}-m_{60}$ color.

\item[\bf Stellar luminosity] In Figs.~\ref{fig:sedpanel}(g) and
  \ref{fig:panel}(g), we show the effect of varying the stellar
  luminosity.  Changing the stellar luminosity influences the
  spectra in two ways.  The inner rim becomes much less pronounced in
  the calculations for a star with $L_{\star}=10L_{\odot}$, an effect
  already noted by DDN01.  At the same time, the low-luminosity object
  contains more flux at long wavelengths, between 60 and 100\um.  In
  particular, this effect leads to a horizontal shift in the diagnostic
  color-color plot.

\item[\bf Carbon/silicate ratio] In Figs.~\ref{fig:sedpanel}(h) and
  \ref{fig:panel}(h), the effect of varying the carbon/silicon mass
  ratio in the dust opacity calculations is shown.  It is immediately
  clear that this effect is minor and hardly affects the general shape
  of the SED.  We will discuss in a companion paper (Meijer et al
  2007, submitted to A\&A) how this is an important parameter in terms
  of the strength of the 10\um{} feature.  However, for the SED shape
  as probed by our diagnostic diagram, the carbon content has little
  or no effect.  SInce the addition of carbon increases the continuum level,
  both the NIR and FIR flux increase at the same rate, so that the
  ratio does not change.  Since the slope between 12~\micron\ and
  60~\micron\ hardly changes, also the IRAS $m_{12}-m_{60}$ color
  remains unchanged.
\end{description}

\section{Discussion}
\label{sec:discussion}

\subsection{Effects of the weak inner rim}
\label{sec:inner-rim-systematics}

\begin{figure}[t]
\centerline{\includegraphics[width=\columnwidth]{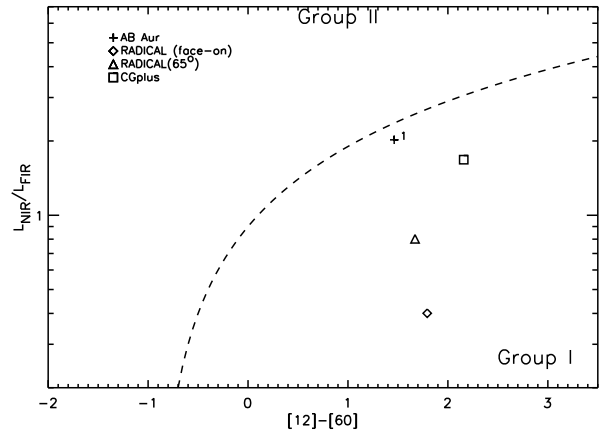}}
\caption[]{\label{fig:abaur-sys} Effect of the strength of the inner
  rim on the position of a source in the color-color diagram.  The
  diagram shows the locations of the measured SED of AB Aurigae, of
  the D03 fit using the DDN model, and of face-on and 65\degree{}
  SEDs of the new model, computed for the same disk parameters.}
\end{figure}

In Sect.~\ref{sec:first-look} we demonstrated that the inner
rim emission in the new models is systematically weaker than in the
simpler DDN models.  In Table \ref{tab:models}, we quantify this
result.  Looking at the flux ratios derived from observed SEDs (see
table \ref{tab:observations}), we can see that the rim emission (the
near-IR flux, listed in Col.\ 10) accounts for the reprocessing of typically
20-30\% of the stellar luminosity.  In the models, the typical value
is 15-17\% and varies only weakly with disk properties.  Clearly,
this should also have a systematic effect on the diagnostic
diagram.  In fig.\ref{fig:abaur-sys}, the infrared colors of AB Aur
are compared with model predictions.  Clearly, the difference is
mainly a shift in the ratio of NIR to FIR luminosity in the disk.
Taking this into account, it is unsurprising that our model
predictions are below the data points of the observations.

\subsection{Effects of moderate dust growth}
\label{sec:effects-moder-dust}

An important part of our parameter study has been considering the effects of
moderate dust growth, i.e. either extending the upper limit of the
size distribution from 1\um to values of up to 25\um, or the complete
removal of small grains, modeled by shifting up the lower boundary of
the size distribution.  In
Sect.~\ref{sec:gener-param-trends} we found that removing grains below
1\um has only a small effect on the structure of the models, while
shifting the upper boundary of the size distribution (and in this way
removing smaller grains as well, including grains with sizes close to
1\um) may affect the global SED.  In this section, we study the basic
effects of changes in grain size  on the strength of the 10\um
feature.  Again, we do this without paying attention to mineralogical
information - only the overall strength of the feature is
considered.

\subsubsection{Weak 10\micron\ feature}
\label{sec:weak-10micr-feat}

Dust emission features in disks are an excellent way of probing the
composition and size of grains present in the disk.  The observed
10\,\um features can vary strongly, both in strength and shape
\citep[e.g.][]{bwd03,2003A&A...412L..43P,bmw05,kad06}.  In particular,
there are a few sources that show no discernible 10\,\um feature at
all.  These objects were classified by \citet{mwb01} into
subclass b.  Observationally, at least in the bright ISO sample
\citep{mwb01}, a significant fraction of objects with SED group Ib
were found, but no clear cases of group IIb.  An important
question is, whether this is an automatic consequence of the disk
flaring geometry (i.e. group I versus group II), or if additional
factors are needed to produce weak features.  We  therefore
measured the feature strength in all our models.

   \begin{figure}[t]
   \centering
     \includegraphics[width=\columnwidth]{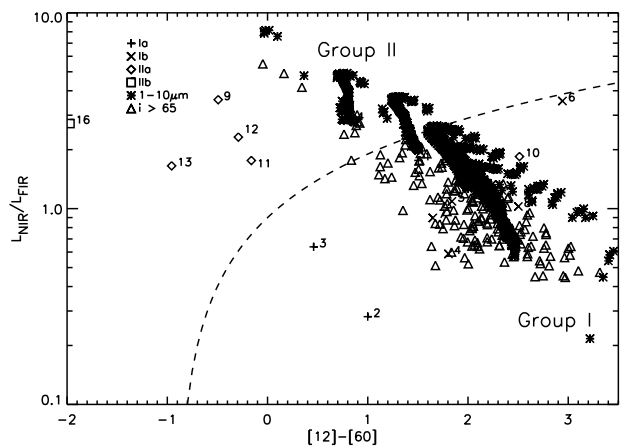}
     \caption{All models in the entire parameter study that formally
       show a weak 10\,\um feature.  We measured the
       peak-to-continuum ratio by drawing a linear continuum between
       \textbf{8} and \textbf{13}\,\um and measuring the peak flux in
       the way described by \citet{bwd03}.  Pluses, crosses, diamonds,
       and squares show the location of observed sources of group Ia,
       Ib, IIa, and IIb, respectively.  Stars indicate models with all
       grains sizes below 1\,\um removed.  Triangles are extreme
       inclinations with weak features caused by self-absorption.}
         \label{fig:nofeature}%
   \end{figure}

\begin{figure}[htbp]
\centerline{\includegraphics[]{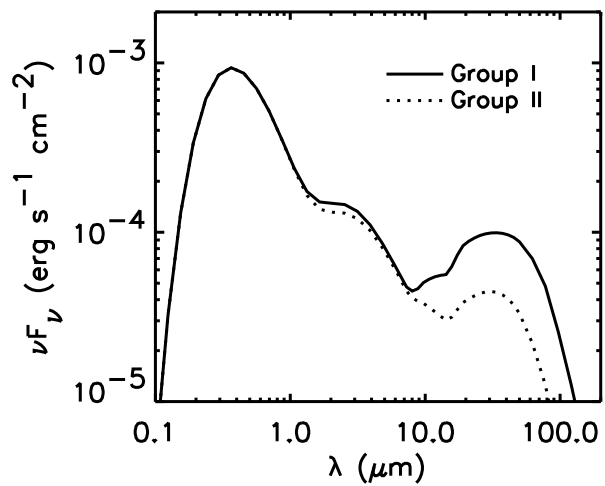}}
\caption[]{\label{fig:sed-weak-feature} Example SEDs of two models
  with weak or absent 10 \um feature.  The group I model has a total
  dust mass of $3\times10^{-5}$\,M$_{\odot}$, the grouop II model has
  $3\times10^{-6}$\,M$_{\odot}$.  The other parameters are the same
  for both models: R$_{\rm out}=200$\,AU, 20\% carbon content,
  $p=-1.5$, and a size distribution for the dust grain of between 1 and
  10\,\um with a slope $m=-3.5$.}
\end{figure}

Most of the models show a strong 10\micron\ feature and would therefore be
classified as group Ia or IIa in the \citet{mwb01} scheme.  However,
there are several corners of parameter space where a weak silicate
feature can exist.  These sources would then have to be classified as
group Ib or IIb.  An overview plot of all models with a
feature-t-continuum ratio lower than 1.05 (see
Fig.~\ref{fig:nofeature}).  There are two basic ways of producing a weak
emission feature:
\begin{enumerate}
\item\label{item:1} \textit{Viewing a disk model at an inclination
    close to but not fully edge-on}.  In this case, the emission
  feature produced by the warm inner disk is partially compensated by
  absorption further out in the disk.  A typical inclination for this
  effect to occur is 65$^{\circ}$.  It is clear that this requires
  fine-tuning of the inclination in the modeling.  Furthermore,
  models with a self-absorbed feature have significantly higher
  reddening.  Observed sources with classification Ib do not exhibit
  higher reddening than group Ia sources.  Therefore, we are left with
  the second possibility:
\item \textit{Efficient removal of grains smaller than about 3\um}.
  In the framework of our model, this can be achieved by either increasing the
  lower limit of the grain size distribution to a few \um{}, or
  increasing the upper limit of the grain size distribution.  If the
  grain size distribution is sufficiently steep (i.e. $m<-3.5$),
  most of the mass resides in large grains, so that increasing the
  upper limit effectively reduces the mass in small grains.  We show
  two example of SEDs without strong features in
  Fig.\ref{fig:sed-weak-feature}. 
\end{enumerate}

While possibility (2) providesdelivers a reasonable explanation for the existence of the
subgroup b sources, an important problem is introduced by the fact
that subgroup b appears to be much more abundant in group I sources than
in group II sources \citep{aa04}, while the models show just as many
sources that could be classified as IIb.  Therefore, we must be
dealing here with either an observational bias, or with an
evolutionary effect.  \citet{aa04} suggested that group IIa sources are
systematically brighter than group IIb sources, so that observational
biases could be at work.

\section{Conclusions}
\label{sec:conclusions}

We arrive at the following conclusions.

\begin{enumerate}
\item\label{item:2} 
\textbf{The Parameter range investigated covers most of observed SEDs}\\
By comparing all models with observations, we are able to explain
most, but we are unable to account for teh observational data points
in the upper right corner for HD 142527 (11), HD 135344 (6) and HD
169142 (8).  These stars all show evidence of disk gaps
in imaging observations, which may be part of the reason why they are
more difficult objects to describe.

\item \textbf{The mass of small grains largely determines the SED type}\\
  We find that the mass in small grains is the single most important
  parameter determining the SED type.  While extreme values of the
  surface density power law (DD04) can also shift a disk from group I
  to II and vice versa, the values required for this to occur seem too
  extreme.  All other parameters have very limited influence on the
  classification of a source as group I or group II.

\item \textbf{No group I/II dichotomy}\\
  From a modeling point of view, we find that there is no clear
  dichotomy between group I and group II models.  The position of a
  model in the diagnostic color-color diagram changes smoothly across
  the dividing line as parameters vary.  The line connecting models of
  different small dust mass is almost perpendicular to the dividing
  line, while changes in other parameters cause shifts that are at
  least partially parallel to the dividing line.  As the dust mass is
  changed in logarithmic steps, there is a gradual transition from
  fully flared (group I) to non-flaring (group II)
  disks.  Some observed sources are in fact close to the dividing
  line, in particular AB Aur (source number 1) and HD 163296 ( source
  number 14) fall into this class.  A scarcity of intermediate
    objects, if confirmed in studies of larger samples, would indicate
    a rapid transition if the distinction is interpreted as an
    evolutionary effect.

\item \textbf{The strength of the inner rim emission is hardly
    dependent on disk mass}

  The emission caused by the inner rim is hardly dependent on model
  parameters, as long as the dominant grain size is the inner rim
  remains the same.  As long as the inner rim is sufficiently optically
  thick, its height depends only weakly on the details.  In
  particular, the small grain mass influences only the inner rim
  emission when the surface density is decreased by more than a factor
  of 100, compared with the case in which all dust mass is in small
  grains.  This is consistent with observations that show the 3\,\um
  bump to be similar in all Herbig stars \citep{mwb01}.

\item \textbf{The 3\um bump in the predicted spectra is too weak}\\
  The models show that with the current assumptions, the
  self-consistent models do still not produce a sufficient amount of
  emission the 3\um bump.  This causes a systematic shift of the
  models in the color-color diagram.  A more detailed implementation
  of the inner rim in models is highly desired.

\end{enumerate}

\begin{acknowledgements}
  J.M. acknowledges financial support from Network\,II of the
  Nederlandse Onderzoeksschool Voor Astronomy (NOVA) through project
  10.10.2.12 and from the NWO Council for Physical Sciences through
  project 614.041.007.
\end{acknowledgements}


\begin{thebibliography}{33}
\expandafter\ifx\csname natexlab\endcsname\relax\def\natexlab#1{#1}\fi

\bibitem[{{Acke} \& {van den Ancker}(2004)}]{aa04}
{Acke}, B. \& {van den Ancker}, M.~E. 2004, \aap, 426, 151

\bibitem[{{Acke} {et~al.}(2004){Acke}, {van den Ancker}, {Dullemond}, {van
  Boekel}, \& {Waters}}]{aad04}
{Acke}, B., {van den Ancker}, M.~E., {Dullemond}, C.~P., {van Boekel}, R., \&
  {Waters}, L.~B.~F.~M. 2004, \aap, 422, 621

\bibitem[{{Bouwman} {et~al.}(2001){Bouwman}, {Meeus}, {de Koter}, {Hony},
  {Dominik}, \& {Waters}}]{bmk01}
{Bouwman}, J., {Meeus}, G., {de Koter}, A., {et~al.} 2001, \aap, 375, 950

\bibitem[{{Chiang} \& {Goldreich}(1999)}]{1999ApJ...519..279C}
{Chiang}, E.~I. \& {Goldreich}, P. 1999, \apj, 519, 279

\bibitem[{{Dohnanyi}(1969)}]{1969JGR....74.2531D}
{Dohnanyi}, J.~W. 1969, \jgr, 74, 2531

\bibitem[{{Dominik} {et~al.}(2003){Dominik}, {Dullemond}, {Waters}, \&
  {Walch}}]{2003A&A...398..607D}
{Dominik}, C., {Dullemond}, C.~P., {Waters}, L.~B.~F.~M., \& {Walch}, S. 2003,
  \aap, 398, 607

\bibitem[{{Dullemond}(2002)}]{d02}
{Dullemond}, C.~P. 2002, \aap, 395, 853

\bibitem[{{Dullemond} \& {Dominik}(2004)}]{dd04}
{Dullemond}, C.~P. \& {Dominik}, C. 2004, \aap, 417, 159

\bibitem[{{Dullemond} \& {Dominik}(2005)}]{2005A&A...434..971D}
{Dullemond}, C.~P. \& {Dominik}, C. 2005, \aap, 434, 971

\bibitem[{{Dullemond} {et~al.}(2001){Dullemond}, {Dominik}, \&
  {Natta}}]{2001ApJ...560..957D}
{Dullemond}, C.~P., {Dominik}, C., \& {Natta}, A. 2001, \apj, 560, 957

\bibitem[{{Dullemond} \& {Natta}(2003)}]{dn03}
{Dullemond}, C.~P. \& {Natta}, A. 2003, \aap, 408, 161

\bibitem[{{Fujiwara} {et~al.}(2006){Fujiwara}, {Honda}, {Kataza}, {Yamashita},
  {Onaka}, {Fukagawa}, {Okamoto}, {Miyata}, {Sako}, {Fujiyoshi}, \&
  {Sakon}}]{2006ApJ...644L.133F}
{Fujiwara}, H., {Honda}, M., {Kataza}, H., {et~al.} 2006, \apjl, 644, L133

\bibitem[{{Fukagawa} {et~al.}(2006){Fukagawa}, {Tamura}, {Itoh}, {Kudo},
  {Imaeda}, {Oasa}, {Hayashi}, \& {Hayashi}}]{2006ApJ...636L.153F}
{Fukagawa}, M., {Tamura}, M., {Itoh}, Y., {et~al.} 2006, \apjl, 636, L153

\bibitem[{{Hillenbrand} {et~al.}(1992){Hillenbrand}, {Strom}, {Vrba}, \&
  {Keene}}]{1992ApJ...397..613H}
{Hillenbrand}, L.~A., {Strom}, S.~E., {Vrba}, F.~J., \& {Keene}, J. 1992, \apj,
  397, 613

\bibitem[{{Isella} \& {Natta}(2005)}]{in05}
{Isella}, A. \& {Natta}, A. 2005, \aap, 438, 899

\bibitem[{{Kamp} \& {Dullemond}(2004)}]{kd04}
{Kamp}, I. \& {Dullemond}, C.~P. 2004, \apj, 615, 991

\bibitem[{{Kessler-Silacci} {et~al.}(2006){Kessler-Silacci}, {Augereau},
  {Dullemond}, {Geers}, {Lahuis}, {Evans}, {van Dishoeck}, {Blake}, {Boogert},
  {Brown}, {J{\o}rgensen}, {Knez}, \& {Pontoppidan}}]{kad06}
{Kessler-Silacci}, J., {Augereau}, J.-C., {Dullemond}, C.~P., {et~al.} 2006,
  \apj, 639, 275

\bibitem[{{Lada}(1987)}]{Lada87}
{Lada}, C.~J. 1987, in IAU Symp. 115: Star Forming Regions, ed. M.~{Peimbert}
  \& J.~{Jugaku}, 1--17

\bibitem[{{Laor} \& {Draine}(1993)}]{ld93}
{Laor}, A. \& {Draine}, B.~T. 1993, \apj, 402, 441

\bibitem[{{Leinert} {et~al.}(2004){Leinert}, {van Boekel}, {Waters},
  {Chesneau}, {Malbet}, {K{\"o}hler}, {Jaffe}, {Ratzka}, {Dutrey}, {Preibisch},
  {Graser}, {Bakker}, {Chagnon}, {Cotton}, {Dominik}, {Dullemond},
  {Glazenborg-Kluttig}, {Glindemann}, {Henning}, {Hofmann}, {de Jong},
  {Lenzen}, {Ligori}, {Lopez}, {Meisner}, {Morel}, {Paresce}, {Pel},
  {Percheron}, {Perrin}, {Przygodda}, {Richichi}, {Sch{\"o}ller}, {Schuller},
  {Stecklum}, {van den Ancker}, {von der L{\"u}he}, \&
  {Weigelt}}]{2004A&A...423..537L}
{Leinert}, C., {van Boekel}, R., {Waters}, L.~B.~F.~M., {et~al.} 2004, \aap,
  423, 537

\bibitem[{{Mathis} {et~al.}(1977){Mathis}, {Rumpl}, \& {Nordsieck}}]{mrn77}
{Mathis}, J.~S., {Rumpl}, W., \& {Nordsieck}, K.~H. 1977, \apj, 217, 425

\bibitem[{{Meeus} {et~al.}(2001){Meeus}, {Waters}, {Bouwman}, {van den Ancker},
  {Waelkens}, \& {Malfait}}]{mwb01}
{Meeus}, G., {Waters}, L.~B.~F.~M., {Bouwman}, J., {et~al.} 2001, \aap, 365,
  476

\bibitem[{{Natta} {et~al.}(2000){Natta}, {Grinin}, \&
  {Mannings}}]{2000prpl.conf..559N}
{Natta}, A., {Grinin}, V., \& {Mannings}, V. 2000, Protostars and Planets IV,
  559

\bibitem[{{Natta} {et~al.}(2001){Natta}, {Prusti}, {Neri}, {Wooden}, {Grinin},
  \& {Mannings}}]{2001A&A...371..186N}
{Natta}, A., {Prusti}, T., {Neri}, R., {et~al.} 2001, \aap, 371, 186

\bibitem[{{Preibisch} {et~al.}(1993){Preibisch}, {Ossenkopf}, {Yorke}, \&
  {Henning}}]{poy93}
{Preibisch}, T., {Ossenkopf}, V., {Yorke}, H.~W., \& {Henning}, T. 1993, \aap,
  279, 577

\bibitem[{{Przygodda} {et~al.}(2003){Przygodda}, {van Boekel},
  {{\`A}brah{\`a}m}, {Melnikov}, {Waters}, \& {Leinert}}]{2003A&A...412L..43P}
{Przygodda}, F., {van Boekel}, R., {{\`A}brah{\`a}m}, P., {et~al.} 2003, \aap,
  412, L43

\bibitem[{{Tanaka} {et~al.}(2005){Tanaka}, {Himeno}, \&
  {Ida}}]{2005ApJ...625..414T}
{Tanaka}, H., {Himeno}, Y., \& {Ida}, S. 2005, \apj, 625, 414

\bibitem[{{van Boekel} {et~al.}(2004){van Boekel}, {Min}, {Leinert}, {Waters},
  {Richichi}, {Chesneau}, {Dominik}, {Jaffe}, {Dutrey}, {Graser}, {Henning},
  {de Jong}, {K{\" o}hler}, {de Koter}, {Lopez}, {Malbet}, {Morel}, {Paresce},
  {Perrin}, {Preibisch}, {Przygodda}, {Sch{\" o}ller}, \& {Wittkowski}}]{bml04}
{van Boekel}, R., {Min}, M., {Leinert}, C., {et~al.} 2004, \nat, 432, 479

\bibitem[{{van Boekel} {et~al.}(2005){van Boekel}, {Min}, {Waters}, {de Koter},
  {Dominik}, {van den Ancker}, \& {Bouwman}}]{bmw05}
{van Boekel}, R., {Min}, M., {Waters}, L.~B.~F.~M., {et~al.} 2005, \aap, 437,
  189

\bibitem[{{van Boekel} {et~al.}(2003){van Boekel}, {Waters}, {Dominik},
  {Bouwman}, {de Koter}, {Dullemond}, \& {Paresce}}]{bwd03}
{van Boekel}, R., {Waters}, L.~B.~F.~M., {Dominik}, C., {et~al.} 2003, \aap,
  400, L21

\bibitem[{{Vinkovi{\'c}} {et~al.}(2006){Vinkovi{\'c}}, {Ivezi{\'c}},
  {Jurki{\'c}}, \& {Elitzur}}]{2006ApJ...636..348V}
{Vinkovi{\'c}}, D., {Ivezi{\'c}}, {\v Z}., {Jurki{\'c}}, T., \& {Elitzur}, M.
  2006, \apj, 636, 348

\bibitem[{{Waters} \& {Waelkens}(1998)}]{ww98}
{Waters}, L.~B.~F.~M. \& {Waelkens}, C. 1998, \araa, 36, 233

\bibitem[{{Whitney} {et~al.}(2003){Whitney}, {Wood}, {Bjorkman}, \&
  {Cohen}}]{2003ApJ...598.1079W}
{Whitney}, B.~A., {Wood}, K., {Bjorkman}, J.~E., \& {Cohen}, M. 2003, \apj,
  598, 1079

\end{thebibliography}
\end{document}